\documentclass[12pt]{elsart}
\usepackage[english]{babel}
\usepackage{graphics}
\renewcommand{\thesection}{\Roman{section}}
\renewcommand{\thetable}{\Roman{table}}
\renewcommand{\thefootnote}{\alph{footnote}}
\newcommand{\be}{\begin{equation}}
\newcommand{\ee}{\end{equation}}
\newcommand{\bc}{\begin{center}}
\newcommand{\ec}{\end{center}}
\newcommand{\bea}{\begin{eqnarray}}
\newcommand{\eea}{\end{eqnarray}}

\def\v{^}
\begin{document}
%
%	First page
%
\begin{frontmatter}
\title{Electronic structure and dimerization of a single monatomic gold wire}
\author{L. De Maria\thanksref{cif}} and
\author{M. Springborg\thanksref{mail}}
\address{Universit\"at Konstanz, Fakult\"at f\"ur Chemie,Universit\"atsstra{\ss}e 10, D-78457 Konstanz, Germany}
\thanks[cif]{Also at Centro Internacional de Fisica Teorica, Apartado Aereo 49490, Santaf\'e de Bogot\'a, Colombia.}
\thanks[mail]{Corresponding author : Michael Springborg, Universit\"at Konstanz, Fakult\"at f\"ur Chemie, Universit\"atsstra{\ss}e 10, Postfach 5560 - M 722, D-78457 Konstanz, Germany; fax: +49 7531 883139; email: {\tt mcs@chclu.chemie.uni-konstanz.de} }

\begin{abstract}
The electronic structure of a single monatomic gold wire is presented
for the first time. It has been obtained with state-of-the-art {\it
ab-initio} full-potential density-functional (DFT) LMTO (linearized
muffin-tin orbital) 
calculations taking into account relativistic
effects. For stretched structures in the experimentally
accessible range the conduction band is exactly half-filled, whereas
the band structures are more complex for the optimized structure. By
studying the total energy as a function of unit-cell length and of a
possible bond-length alternation we find that the system can lower its
total energy by letting the bond lengths alternate leading to a
structure containing separated dimers with bond lengths of about 2.5 \AA, 
largely independent of the stretching. However, first for fairly large
unit cells (above roughly 7 \AA), is the total-energy gain upon this
dimerization comparable with the energy costs upon stretching. We propose
that this together with band-structure effects 
is the reason for the larger interatomic distances observed 
in recent experiments. We find also that although spin-orbit couplings 
lead to significant effects on the band structure,
the overall conclusions are not altered, and that finite Au$_2$, Au$_4$, and
Au$_6$ chains possess electronic properties very similar to those of the
infinite chain.
\end{abstract}
\end{frontmatter}
\newpage

\normalsize

Very recently, single chains of suspended gold atoms have been
produced between two [110]-oriented tips in
transmission electron microscope 
(TEM) \cite{exp1} and in mechanically controllable
break-junction (MCB) experiments \cite{exp2}. In the TEM experiment,
conductance measurements were performed simultaneously with 
electron-microscope images of the atomic-size contact as the tips were
separated. Before breaking, the contact consists of a monatomic nanowire
made up of four gold atoms and shows a conductance of
$2e^2/h$. The MCB experiment, although lacking direct imaging of
the systems, gives also evidence of the formation of a
monatomic wire upon stretching.

The formation of such a monatomic chain structure for gold and its 
properties upon stretching were studied 
theoretically by S\o rensen $et$ $al.$ \cite{simu} by 
means of classical molecular-dynamics 
simulations although for contacts between [100]-oriented tips. More recently,
Torres $et$ $al.$ \cite{tosatti} studied the thermodynamical
foundations of the spontaneous thinning process as well as the 
stability of the monatomic gold wire for the [110]-oriented tips 
using classical many-body force
simulations as well as {\it ab-initio} local-density (LDA) and 
generalized-gradient (GGA) density-functional electronic-structure calculations.
They found that for bond lengths above 2.8 \AA, 
the system will break into isolated Au$_2$ dimers. However, in the 
experimental studies, the finite
monatomic wire shows an overwhelming stability upon stretching and is 
stable for bond lengths up to between $3$ and $4$ \AA\ .

In the present work the problem of the stability of a single monatomic
gold wire is addressed. The electronic structure of the system is
presented for the first time and its behaviour upon 
stretching used to rationalize the whole scenario. In
particular we show that, upon stretching, the ordering of the
energy levels leads to a situation with one exactly 
half-filled electronic band so
that dimerization (i.e., bond-length alternation) becomes
favoured. We show how, as the system is
stretched, the energy gain upon dimerization increases, but the
nearest-neighbour interatomic distance at equilibrium stays
constant at approximately $2.5$ \AA. Finally, the importance of
relativistic effects, not considered in any previous study, is addressed
and the results for the infinite chains are compared with similar ones for
finite Au$_2$, Au$_4$, and Au$_6$ chains.

We performed full-potential density-functional 
LMTO (linearized muffin-tin orbital) 
calculations with a local-density approximation (LDA) on
isolated periodic infinite gold chains using the method described in 
Refs. \cite{michael1,michael2}. This method is specifically targeted for
isolated, infinite, periodic, helical polymers and chain compounds
and have been applied 
successfully to a wide series of systems \cite{michael3}. The basis set
consists of two sets of $s, p$ and $d$ functions on all sites; each
function is defined numerically inside non-overlapping atom-centered
spheres and in terms of spherical Hankel functions in the interstitial
region. The two sets differ mainly in decay constants of the latter.
Scalar relativistic (SR) \cite{karla} corrections were included in
{\it all} the calculations presented here; in addition the effects of
spin-orbit (SO) couplings were also considered.

For the calculations on the isolated, undimerized 
gold chain we assume that the nuclei are
lying in the $(x,z)$ plane with the $z$ axis parallel to the chain
axis. We use 31 equidistant
points in half-part of the first Brillouin zone
in order to ensure the appropriate convergence of all the physically
relevant quantities and to properly describe the metallicity of the
system. We also use the level-broadening scheme described in
\cite{broad} with an electronic temperature of $0.068$ eV
($5\times10^{-3}$ Ry). For the dimerized chain we use 16 points
in half-part of the first Brillouin zone. 

The structure of the undimerized chain can be described with the single 
bond length $a$, whereas the dimerized chain has two alternating bond
lengths $a_1$ and $a_2$. From these we define an average bond length
\begin{equation}
\bar a={1\over2}(a_1+a_2)
\label{avera}
\end{equation}
and a dimerization coordinate
\begin{equation}
\delta={1\over2}(a_1-a_2).
\label{dimco}
\end{equation}

The electronic band structures of a single monatomic gold wire with an
interatomic distance of $a=$3.5 \AA\ are shown in Fig.\ \ref{fig:ele}.a.
This bond length is representative of those observed experimentally. 
The system is metallic with a half-filled band of symmetry $\sigma$.
Below this, a broader occupied $\sigma$ band as well as (doubly 
degenerate) narrower $\pi$ and $\delta$ bands are found. Analyzing the
orbitals it turns out that the $\sigma$ bands have contributions from
both $6s$ and $5d_{z\v2}$ functions, 
whereas the $\pi$ and $\delta$ bands largely are due to $d$ functions.

Including the spin-orbit (SO) couplings leads to the band structures of
Fig.\ \ref{fig:ele}.b. Except for the fact that the lower symmetry leads to
a splitting of the doubly degenerate bands as well as to avoided crossings
between various pairs of bands, the overall picture is not altered and, most
important, the occurrence of one exactly half-filled band remains.

Finally, Fig.\ \ref{fig:ele}.c shows the energy levels for finite
Au$_N$ chains consisting of $N=2$, $4$, and $6$ atoms, respectively. In these 
calculations we set all bond lengths equal to 3.5 \AA, but we stress that
these do $not$ correspond to the optimized values (for the dimer, the
optimized bond length is in fact much shorter, as we shall see below).
Instead, they are similar to the ones observed experimentally.

In Fig.\ \ref{fig:ele}.c it can be seen that
although increasing the number of atoms leads to some broadening of the
energy regions spanned by the orbitals,
most of the features of the infinite systems are recovered already
for these fairly small systems. Most notably, the fact that $\sigma$ 
orbitals are those appearing closest to the Fermi level and that $\pi$ and
$\delta$ orbitals appear at deeper energies is true also for the finite
chains. Furthermore, 
the Fermi energies of the finite systems are very similar to those
of the infinite chains.

In Fig.\ \ref{fig:coh} the total energy of the nanowire is shown as a
function of the interatomic distance $a$. Empty squares (triangles)
indicate values of the cohesive energy for which the relativistic contributions
were included up to the SR (SO) level.
At the SR level, an equilibrium distance of about $2.65$
\AA\ is predicted together with a total-energy minimum of $1.35$ eV/atom
in good agreement with the results of the other {\it ab-initio} calculations
\cite{tosatti}. The inclusion of the SO coupling leads to a
reduction of the equilibrium length to $2.55$ \AA\ and an increase of
the cohesive energy by approximately $0.10$ eV/atom. Such contractions due
to relativistic effects are often observed (see, e.g., \cite{rela}).

In both curves of Fig.\ \ref{fig:coh} the energy values between $2.7$
and $2.9$ \AA\ are not shown. For these, numerical problems obscured
the calculations, i.e., the highest occupied $\pi$ band for $k=0$ 
(cf. Fig.\ \ref{fig:ele}.a) was lifted to so high energies that it became
partly empty. This placed the Fermi level very close to a van Hove
singularity in the density of states leading to smaller discontinuities in
the total-energy curve. Although the effects were very tiny they were
observable. In addition, they show how sensible the system is
to {\it external} perturbations. Below 2.7 \AA\ the doubly degenerate
$\pi$ band at $k=0$ in Fig.\ \ref{fig:ele}.a is lifted even further so that
this band is only partially filled and the broad $\sigma$ band no longer
is exactly half-filled.

An exactly half-filled band, as found for $a$ above 2.9 \AA, favours 
strongly a (Peierls) dimerization. Therefore, in 
Fig.\ \ref{fig:dim}
the energy gain upon dimerization is shown as a
function of the dimerization coordinate $\delta$ for some selected 
values of the average bond length $\bar a$ that lie 
in the experimentally accessible region. In contrast to previous
calculations \cite{tosatti}, we find that the system possess a stable
structure consisting of alternating shorter and longer bonds where the
shorter bonds have lengths of about 2.5 \AA. Furthermore, by comparing with
the total-energy curves of Fig.\ \ref{fig:coh} we see that the energy gain
upon dimerization first for $\bar a$ about 3.5 \AA\ becomes 
comparable with the energy costs upon stretching. 

Trans polyacetylene
(CH)$_x$ is a well-known example of a material that possesses a Peierls
dimerization, leading to alternating single and double bonds between the
carbon atoms of the backbone. For this, however, 
both the amplitude of the bond-length
alternation as well as the related total-energy gain are much smaller than
those observed here for the gold chain \cite{michael2}.

In Fig.\ \ref{fig:dimele} we show the band structures for two representative
values of $\bar a$, and for each case we show them for two values of the
dimerization coordinate $\delta$, i.e., a very small one and the optimized 
value in Fig.\ \ref{fig:coh}. The occurrence of a band gap at
the Fermi level due to the dimerization is readily recognized for the smaller
value of $\delta$. For larger values of $\delta$, at least for the larger
values of $\bar a$, the highest occupied orbital is no longer derived from
the $\sigma$ band crossing the Fermi level for the undimerized structure, but
from bands of $\pi$ or $\delta$ symmetry.
Moreover, for these larger values of $\bar a$, the bands
become fairly flat for the optimized structure 
which indicates that at those distances the electronic
interactions between the dimers are only weak. For the sake of comparison
we show in the figure also the single-particle energies for the isolated
dimer with a bond length of 2.5 \AA. These are seen to lie fairly close to
the band regions for the optimized structures supporting that this 
structure essentially consists of weakly interacting Au$_2$ units.

The fact that the nature of the band gap changes from being a direct
gap between the two $\sigma$ bands to becoming indirect for larger values
of $\delta$ is seen in Fig.\ \ref{fig:gap} that shows the band gap as
a function of $\delta$ for different larger values of $\bar a$. 
For smaller values of $\bar a$, the above-mentioned fact that the $\sigma$
band no longer is exactly half-filled and that the $\pi$ bands are partly
emptied makes the smallest band gap vanishing up till some $\bar a$-dependent
threshold for $\delta$. 

It is remarkable that for
the smallest values of the dimerization coordinate $\delta$ all curves
lie on top of each other. Assuming that the $\sigma$ band crossing the
Fermi level for the undimerized structure can be described with a single
Wannier function per atom and that only nearest-neighbour hopping integrals
need to be taken into account, the band gap is 
\begin{equation}
E_{\rm gap}=2\vert t_1-t_2\vert, 
\label{gaphop}
\end{equation}
where $t_1$ and $t_2$ are the two hopping integrals. These may in turn be 
assumed to depend linearly on the bond lengths,
\begin{equation}
t_{1,2}=t_0-\alpha\cdot(a_{1,2}-\bar a).
\label{hop}
\end{equation}
$\alpha$ is an
electron-phonon coupling constant. For any average bond length $\bar a$
one would expect both $t_0$ and $\alpha$ to depend on $\bar a$. However, since
Eqs.\ (\ref{gaphop}) and (\ref{hop}) imply that
\begin{equation}
E_{\rm gap}=4\alpha \vert\delta\vert,
\label{gap}
\end{equation}
we obtain that the electron-phonon coupling constant is independent of $\bar a$, 
at least in the range considered here. Since the 
tendency towards dimerization largely is determined by the size of the 
electron-phonon coupling, this result implies that the strength of this
tendency is independent of the unit-cell length. Finally, the fact that the
curves of Fig.\ \ref{fig:gap} for larger $\delta$ 
do not lie on the top of each other is due to
the above-mentioned change of the nature of the gap. 

Experimentally, it is found that the finite chains stay stable up to 
bond lengths of about 4 \AA, after which the chains break. Compared with
the systems we have studied here there is a number of differences. First, the
experimentally studied systems consist of only about four atoms, but
as Fig.\ \ref{fig:ele} shows, already such systems possess electronic 
properties close to those of the infinite chain. Second, 
the experimental systems are not isolated but suspended between two tips,
which may give further support for studying infinite chains. Third, and
more importantly, the experimental systems are not static but subjected both
to mechanical (stretching) forces and to electrostatic forces (due to the
voltage between the tips). Here, we have only considered the static parts
of the mechanical forces, which, however, due to the very different time
scales between structural relaxations and the applied forces is justified.
On the other hand, the applied electrostatic forces may influence the 
physically properties since they lead to an overall asymmetric potential
along the chain, although the potential is weak (in the 10 meV range).

In total, our results suggest the following for
the experimental systems. For a given structure with a certain average 
bond length $\bar a$ 
obtained through stretching, the system may attempt to lower
its total energy upon structural relaxation. Here, our results show that
the overall driving mode for this is to split the system into more or less
strongly interacting dimers, although band-structure effects give that this
happens first for $\bar a$ above around 3 \AA. As a competition to this, the system may
attempt to relax towards the shorter, optimized, undimerized structure. Although this
latter is prohibited by the external mechanical forces, we may
still compare the two relaxation modes and, then, first for average bond
lengths above about 3.5 \AA\, is the former energetically preferred. 
Furthermore, due to the external voltage, the dimerization mode is
weakened. Therefore, we suggest that first when $\bar a$ is so large that the
energy gain upon the two relaxation modes are comparable, will the system
change structure, i.e., split into fairly well separated dimers. This offers
thus an explanation for
the unusually long average bond lengths that are observed experimentally.

In conclusion we have shown here how it is necessary to analyze the
electronic band structures of a single monatomic gold chain in order to
gain more understanding into the puzzling problem of its stability. In
particular, we have shown that the dimerization is the most relevant
structural relaxation that shall be considered for these chains.
Furthermore, we
have shown that relativistic effects have significant effects on the
band structures although, maybe surprisingly, without changing the general
picture. To our knowledge, relativistic effects have not been considered
previously for such systems. Finally, we demonstrated that the finite chains
have properties very similar to those of the infinite chain.

The authors want to thank Dr. Karla Schmidt for very useful comments
about the relativistic corrections
and valuable help and guidance concerning the use of the
programs. This project was supported by the Deutsche
Forschungsgemeinschaft (DFG) through project No.\ Sp439/6--1. Finally, the
authors are grateful to Fonds der Chemischen Industrie for very generous
support.

\newpage

%\newpage
\pagebreak

\begin{figure}
\caption{(a-b) Electronic energy bands of an infinite, monatomic gold 
wire and (c) electronic energy levels of some finite linear Au$_N$ 
chains. The interatomic distance for all calculations
is set equal to 3.5 \AA. 
In (c), $+$, $\circ$, and $\times$ mark $\pi$ and
$\delta$ orbitals, empty $\sigma$ orbitals, and filled $\sigma$ orbitals,
respectively. In (a-b), $k=0$ and $k=1$ represent the center and edge of the
first Brillouin zone, respectively, 
and the vertical dashed lines the Fermi level.}
\label{fig:ele} 
\end{figure}

\begin{figure}
\caption{The cohesive energy in eV per atom for a monatomic
gold wire with non-alternating bond lengths. Empty
squares (triangles) represent results obtained
including relativistic contributions up to the SR (SO) level. Lines
are drawn as a guide-to-the-eye.}
\label{fig:coh} 
\end{figure}

\begin{figure}
\caption{Energy gain upon dimerization for selected unit-cell lengths.
The results are shown as a function of the 
dimerization coordinate $\delta$,
and the different curves are obtained for the different average Au--Au
distances $\bar a$ shown in the figure.}
\label{fig:dim} 
\end{figure}

\begin{figure}
\caption{Band structures for infinite chains with alternating bond 
lengths The extra symbols in (b) and (d) represent the single-particle 
energies for an isolated Au dimer with a bond length of 2.5 \AA\ and using
a notation as that of Fig.\ \ref{fig:ele}. The values of $(\bar a,\delta)$
in the four panels are (a) (3.0,0.1), (b) (3.0,0.4), (c) (3.6,0.1), (d)
(3.6,1.1) in \AA.}
\label{fig:dimele}
\end{figure}

\begin{figure}
\caption{The smallest band gap between occupied and unoccupied orbitals
as a function of the dimerization coordinate $\delta$ for the same
values of $\bar a$ and using the same notation as in Fig.\ \ref{fig:dim}.
The horizontal dashed line is the value for the isolated Au$_2$ dimer
with a bond length of 2.5 \AA.}
\label{fig:gap}
\end{figure}


\begin{thebibliography}{99}

\bibitem{exp1} H. Ohnishi, Y. Kondo, and K. Takayanagi, Nature {\bf
395} (1998) 780.

\bibitem{exp2} A.I. Yanson, G. Rubio Bollinger, H.E. van der Brom,
N. Agra\"{\i}t, and J.M. van Ruitenbeek, Nature {\bf 395} (1998) 783.

\bibitem{simu} M.R. S{\o}rensen, M. Brandbyge, and K.W. Jacobsen,
Phys. Rev. B {\bf 57} (1998) 3283.

\bibitem{tosatti} J. A. Torres, E. Tosatti, A. Dal Corso,
F. Ercolessi, J. J. Kohanoff, F. D. Di Tolla, and  J. M. Soler,
cond-mat/9812369 .

\bibitem{michael1} M. Springborg and O.K. Andersen, J. Chem. Phys. {\bf
87} (1987) 7125.

\bibitem{michael2} M. Springborg, J.L. Calais, O. Goscinski, and
L.A. Eriksson, Phys. Rev. B {\bf 44} (1991) 12713.

\bibitem{michael3} M. Springborg, in: M. Springborg (Ed.),
Density-Functional Methods in Chemistry and Materials Science, John
Wiley \& Sons, Chichester, 1997, p. 207 and references therein.

\bibitem{karla} Calculation at the scalar-relativistic level included
the so-called mass-velocity correction and the Darwin term.

\bibitem{broad} M. Springborg, R.C. Albers, and K. Schmidt,
Phys. Rev. B {\bf 57} (1998) 1427.

\bibitem{rela} P. Pyykk{\"o}, Chem. Rev. {\bf 88} (1988) 563. 


\end{thebibliography}
\end{document}